\documentclass[prb,twocolumn,a4paper,superscriptaddress,showpacs,floatfix]{revtex4-1}

\usepackage{dcolumn,amsmath,xspace}
\PassOptionsToPackage{caption=false}{subfig}
\usepackage{subfig}
\usepackage{graphicx}
\usepackage{multirow}
\usepackage{dcolumn}
\usepackage{amssymb}
\usepackage{latexsym}

\usepackage[utf8]{inputenc}
\usepackage{lineno}

\usepackage{epstopdf}

\begin{document}

\title{Interface Electronic Structure in a Metal/Ferroelectric Heterostructure under Applied Bias}

\author{J. E. Rault}
\affiliation{CEA, DSM/IRAMIS/SPCSI, F-91191 Gif-sur-Yvette Cedex, France}
\author{G. Agnus}
\affiliation{Institut d’Electronique Fondamentale, Université Paris-Sud, Bâtiment 220, 91405 Orsay, France}
\author{T. Maroutian}
\affiliation{Institut d’Electronique Fondamentale, Université Paris-Sud, Bâtiment 220, 91405 Orsay, France}
\author{V. Pillard}
\affiliation{Institut d’Electronique Fondamentale, Université Paris-Sud, Bâtiment 220, 91405 Orsay, France}
\author{Ph. Lecoeur}
\affiliation{Institut d’Electronique Fondamentale, Université Paris-Sud, Bâtiment 220, 91405 Orsay, France}
\author{G. Niu}
\affiliation{Université de Lyon, Ecole Centrale de Lyon, Institut des Nanotechnologies de Lyon, F-69134 Ecully cedex, France}
\author{B. Vilquin}
\affiliation{Université de Lyon, Ecole Centrale de Lyon, Institut des Nanotechnologies de Lyon, F-69134 Ecully cedex, France}
\author{M. G. Silly}
\affiliation{Synchrotron-SOLEIL, BP 48, Saint-Aubin, F91192 Gif sur Yvette CEDEX, France}
\author{A. Bendounan}
\affiliation{Synchrotron-SOLEIL, BP 48, Saint-Aubin, F91192 Gif sur Yvette CEDEX, France}
\author{F. Sirotti}
\affiliation{Synchrotron-SOLEIL, BP 48, Saint-Aubin, F91192 Gif sur Yvette CEDEX, France}
\author{N. Barrett}
\email[Correspondence should be addressed to ]{nick.barrett@cea.fr}
\affiliation{CEA, DSM/IRAMIS/SPCSI, F-91191 Gif-sur-Yvette Cedex, France}

\begin{abstract}
The effective barrier height between an electrode and a ferroelectric (FE) depends on both macroscopic electrical properties and microscopic chemical and electronic structure. The behavior of a prototypical electrode/FE/electrode structure, Pt/BaTiO$_\mathrm{3}$/Nb-doped~SrTiO$_\mathrm{3}$, under in-situ bias voltage is investigated using X-Ray Photoelectron Spectroscopy. The full band alignment is measured and is supported by transport measurements. Barrier heights depend on interface chemistry and on the FE polarization. A differential response of the core levels to applied bias as a function of the polarization state is observed, consistent with Callen charge variations near the interface.
\end{abstract}
\vspace*{4ex}

\maketitle
\newpage


\section{Introduction}

The defining property of a Ferroelectric (FE) material is a spontaneous macroscopic polarization which can be reversed under an applied electric field. This has attracted wide interest, the perspective of strain-engineering films for FE-based electronics \cite{pan_enhancement_2004}. Switching the polarization of such films requires a metallic contact, raising fundamental issues on the behavior of the interface between the FE layer and the electrode. The polarization leads to fixed charge of opposite sign at the two metal-FE interfaces. Free charge carriers in the metal electrodes act to screen the polarization charge creating dipoles of the same sign arise at the two interfaces. This screening is usually imperfect, the residual depolarizing field inside the FE alters the electrostatic potential~\cite{junquera_critical_2003, stengel_enhancement_2009, sai_ferroelectricity_2005}. Furthermore, the partially covalent nature of the bonds in the FE changes the band structure with respect to that of a perfectly ionic compound~\cite{stengel_band_2011}. Intense theoretical activity has been deployed to understand the electronic structure and band alignment at the interface~\cite{junquera_critical_2003, nunez_onset_2008, stengel_band_2011, nunez_interface_2008, sai_ferroelectricity_2005, stengel_enhancement_2009, bilc_electroresistance_2012}. First principles calculations can even predict an ohmic barrier but as Stengel et al. point out, this may be due to the underestimation of the band gap in the local density approximation~\cite{stengel_band_2011}. Pintilie \textit{et al.} have extended semiconductor theory of the metal/insulator interface to the case of FE capacitors, including the effect of the polarization on the band lineup and consequential transport properties~\cite{pintilie_ferroelectric_2007}. Umeno et al. predict that the polarization state of FE capacitor can change the barrier height by up to 1~eV~\cite{umeno_ab_2009}.

Despite these important theoretical advances, there is little direct experimental data, due to the intrinsic difficulty of measuring the electronic structure of a buried interface. Such measurements under applied bias are even more challenging. Several groups have conducted electrical measurement on these systems~\cite{pintilie_ferroelectric_2007, khassaf_potential_2012, tyunina_dielectric_2011} but they do not directly probe the microscopic interfacial electronic structure. X-ray Photoelectron Spectroscopy (XPS) can reveal the electronic structure. It has been used to measure an electron barrier height of 0.5~eV at the Pt/BaSrTiO$_\mathrm{3}$ (BSTO) ~\cite{schafranek_barrier_2008} and the role of oxygen vacancies was discussed. The polar dependence interface dipole of PbTiO$_\mathrm{3}$/LaSrMnO$_\mathrm{3}$ (PTO/LSMO) was studied with XPS~\cite{wu_direct_2011}. Zenkevich \textit{et al.} have measured a band offset of 1.42~eV at the top interface of a Pt/BaTiO$_\mathrm{3}$/Cr tunnel junction with photoemission spectroscopy and used electrical measurement to deduce a barrier height change of 0.45~eV when switching ferroelectric polarization~\cite{zenkevich_electronic_2013}. Chen and Klein used XPS with \textit{in-situ} bias to probe the interface between single crystal BaTiO$_\mathrm{3}$ (BTO) and Pt and RuO$_\mathrm{2}$ electrodes~\cite{chen_polarization_2012}. They reported a reversible rigid band shift and barrier height change of 0.6~eV (1.1~eV) for the Pt/BTO (RuO$_\mathrm{2}$/BTO) interface when switching the ferroelectric polarization. 

In this article we study the polarization dependent band alignment and electronic structure of a metal/FE interface using XPS with in-situ biasing. The band alignment is determined by a combination of imperfect screening by the electrode and the chemistry of the interface. We demonstrate that a complete Pt/BTO/NSTO heterostructure has a rectifying Schottky behavior. The band alignment agrees well with the observed transport properties. We suggest that Callen dynamical charge offers a plausible explanation of the polarization dependent electronic structure near the electrode.

\section{Experiment}
\label{sec:experiment}

BaTiO$_\mathrm{3}$ (BTO) thin film was grown on a Nb-doped (0.5~wt.\%) SrTiO$_\mathrm{3}$ (001) (NSTO) conducting substrate using molecular beam epitaxy. Prior to film growth, the NSTO substrate was cleaned using buffered HF solution and rinsed in deionized H$_\mathrm{2}$O, followed by annealing under oxygen atmosphere in the MBE chamber to obtain a clean, atomically flat NSTO surface. Oxygen was introduced into the reactor via a pressure-regulated plasma chamber which enables precise control of oxygen partial pressure and the utilization of atomic (O) oxygen which gives better ferroelectric properties~\cite{niu_molecular_2012}. Ba and Ti were co-evaporated using Knudsen cells at 620~$^{\circ}$C under a oxygen pressure of 2.67x10$^\mathrm{-4}$~Pa and the sample was cooled down under P$_\mathrm{O_2}$ = 1.33x10$^\mathrm{-4}$~Pa. We used Reflection High Energy Electron Diffraction (RHEED) to monitor the film crystallinity during the growth and guarantee the layer by layer growth and the TiO$_\mathrm{2}$ termination. A X-ray diffractometer with a 1.6 kW fixed anode (Cu K$\alpha$ radiation, $\lambda = 1.5406~\AA$) is used to measure the BTO in and out-of-plane parameters as shown in Fig.~\ref{fig:carac_Diff}.
After 10 minutes exposure to ozone to remove the surface carbon contamination, a continuous 2.8-nm Platinum (Pt) top electrode was deposited at room temperature to ensure full coverage of the surface~\cite{schafranek_barrier_2008}. X-ray Photoemission Spectroscopy (XPS) was performed before and after platinum deposition using monochromatized X-Ray Source XM 1000 MkII (Al K$\alpha$: h$\nu$ = 1486.7~eV) and a SPHERA II analyzer, both from Omicron Nanotechnology GmbH.

\begin{figure}[h]
  \centering
   	\subfloat{\label{fig:dispo_scheme}\includegraphics[scale=0.46]{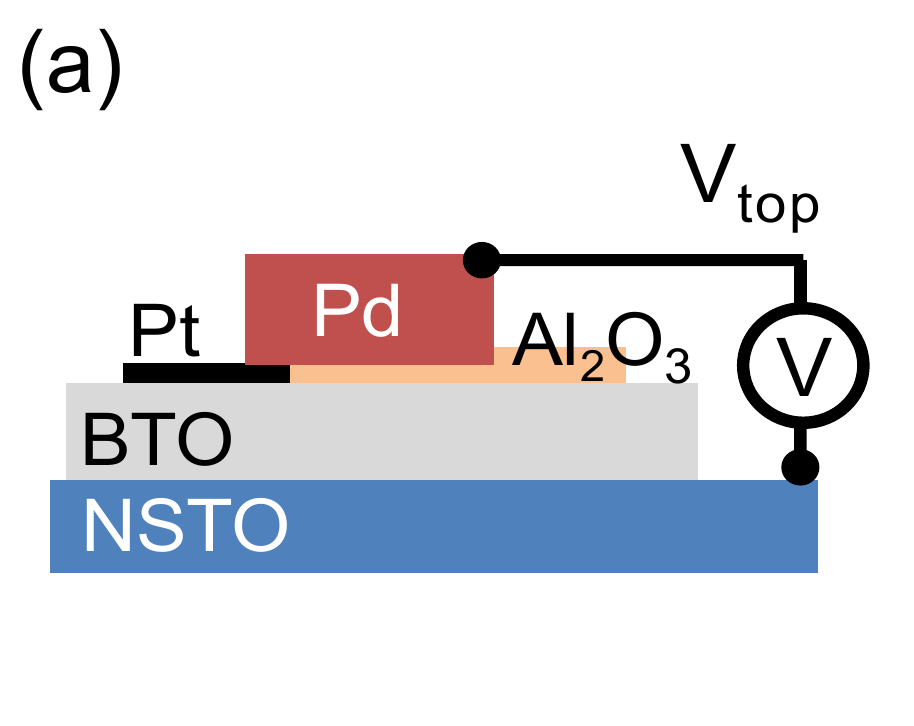}}\hfill
	\subfloat{\label{fig:map_Pt}\includegraphics[scale=0.46]{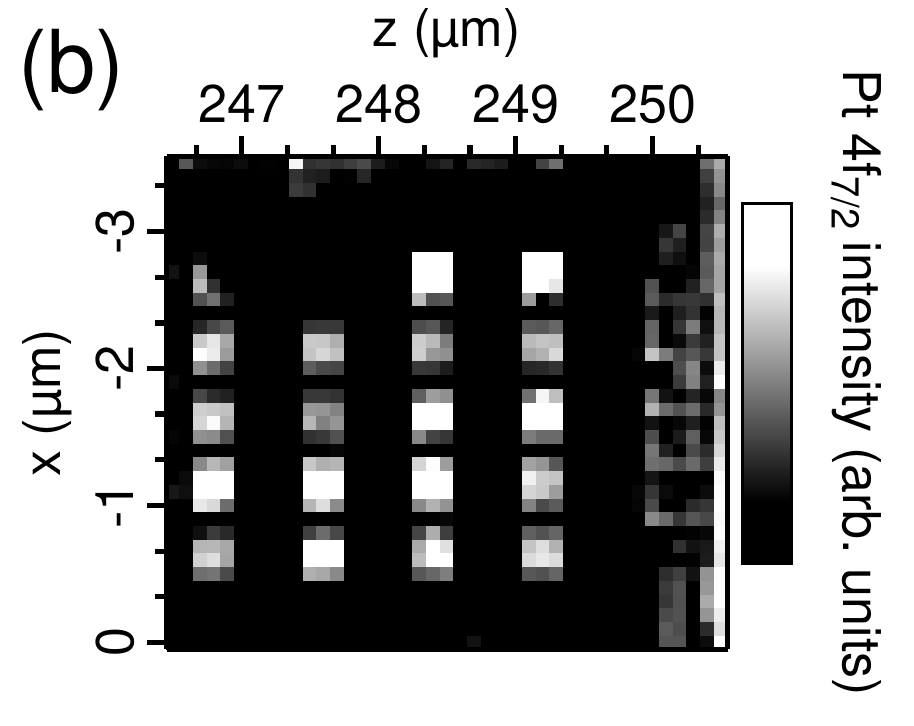}}
  \caption{(a) Schematic of the capacitor; (b) Pt~4f intensity map for the Pt/BTO/NSTO sample showing 20 identical Pt/BTO/NSTO capacitors (300x300 $\mu \mathrm{m}^2$) on the 5x5 mm$^2$ surface, allowing location of the wired capacitor.}
  \label{fig:dispositif}
\end{figure}

300x300~$\mu$m$^\mathrm{2}$, 3 nm thick Pt electrodes were patterned by reactive ionic etching. Thicker (300 nm) Palladium (Pd) pads overlapping part of the Pt electrodes have been deposited by evaporation to enable wire-bonding of the top electrodes to the sample holder. A highly insulating layer of Al$_\mathrm{2}$O$_\mathrm{3}$ was deposited by evaporation onto bare BTO to suppress interference of the Pd pads with the capacitance (see Fig.~\ref{fig:dispo_scheme}). The full device architecture is shown in Fig.~\ref{fig:dispo_scheme}).
The device was then introduced in ultrahigh vacuum (10$^\mathrm{-8}$~Pa) in the XPS set-up of the TEMPO Beamline at the SOLEIL synchrotron radiation source~\cite{polack_tempo:_2010}. The 100x100~$\mu$m$^\mathrm{2}$ beam could be directed onto a single top electrode located by a map of the whole sample using the Pt absorption edge (see Fig.~\ref{fig:map_Pt}). A photon energy of 1100~eV was used to optimize the signal from the BTO close to the interface (estimated probing depth of $\approx$5~nm). This energy is close to those used in typical laboratory based experiments using thin top electrodes~\cite{chen_polarization_2012} but, thanks to synchrotron radiation, the flux has a much higher brilliance and better energy resolution. On the other hand, the higher photon flux can also modify the switching behavior of the capacitor, as will be shown. The overall energy resolution was 220~meV. The sample-holder allows in-situ bias application and electrical measurements via high quality coax wires to limit parasitic behavior due to the electrical environment.

\section{Results}

	\subsection{Pt/BTO/NSTO growth}

\begin{figure}[ht]
  \centering
 		\subfloat{\label{fig:XRD_c}\includegraphics[scale=0.34]{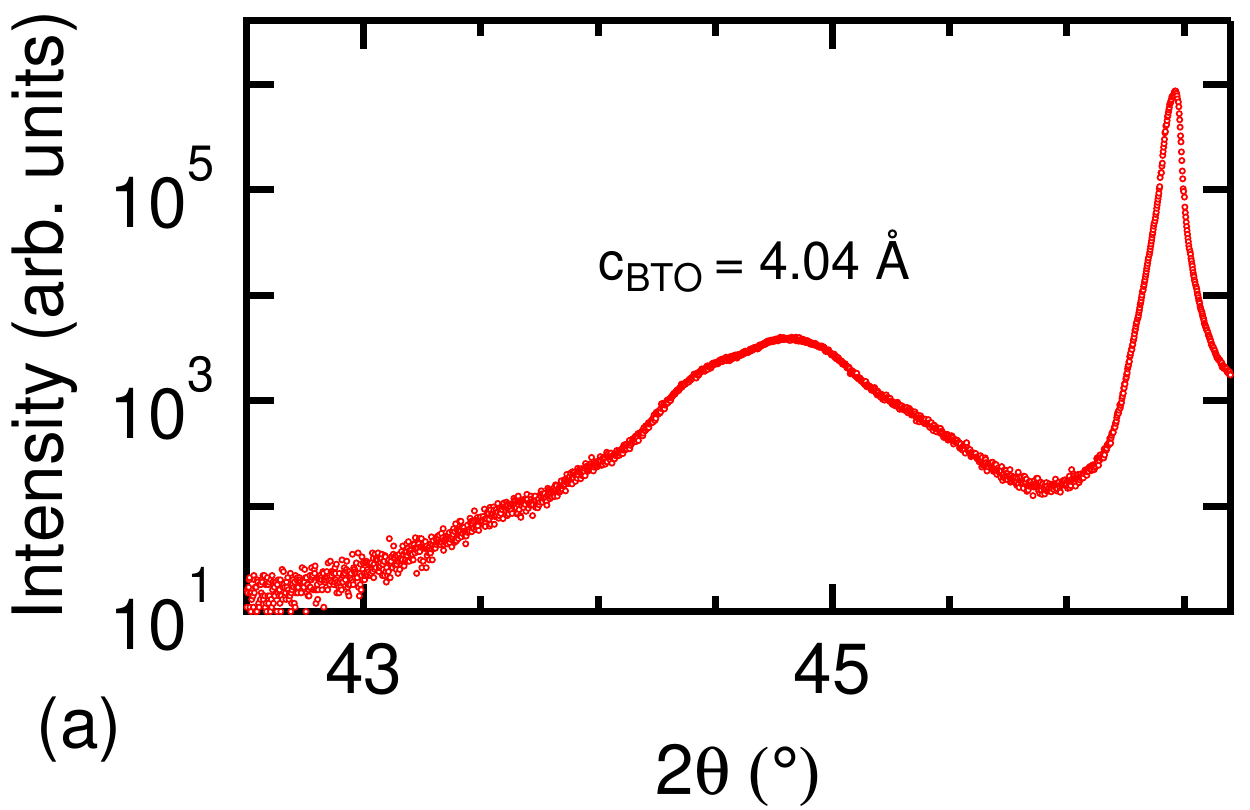}}\hfill
 		\subfloat{\label{fig:XRD_a}\includegraphics[scale=0.34]{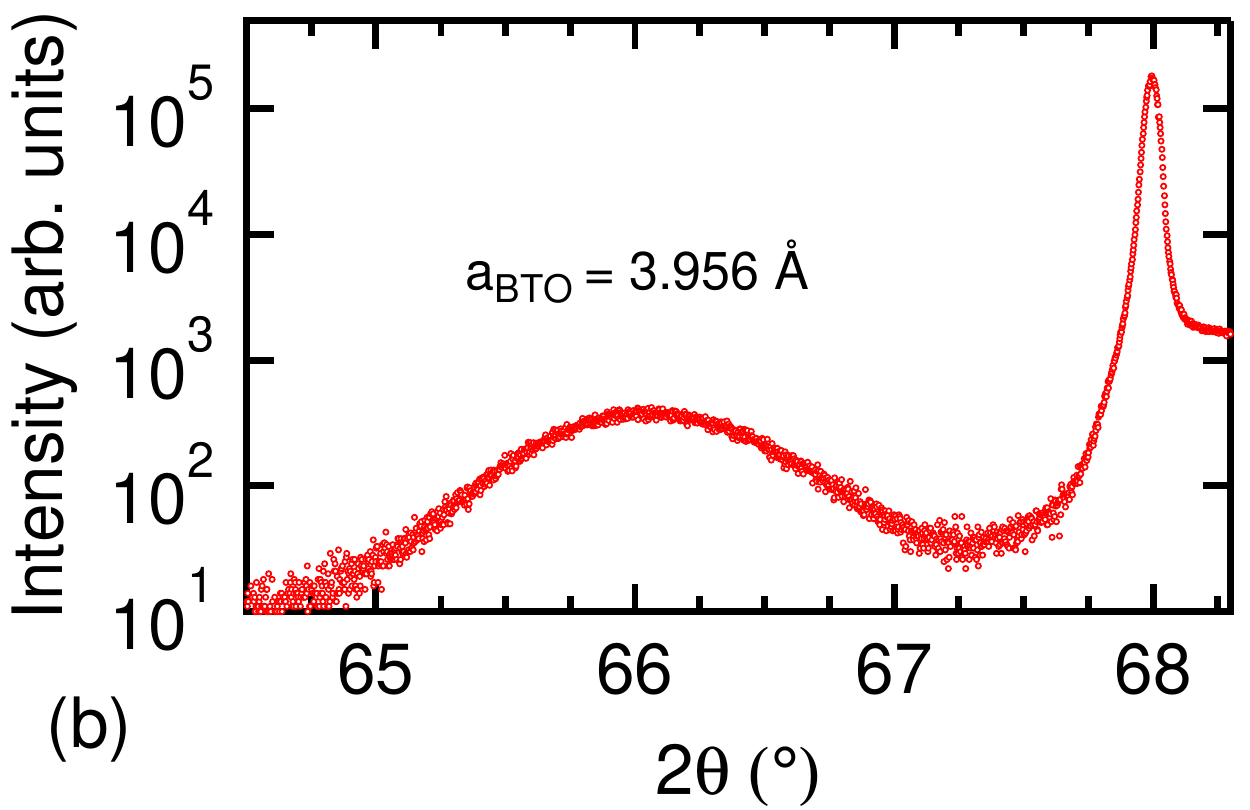}}\\
 		\subfloat{\label{fig:XRR}\includegraphics[scale=0.34]{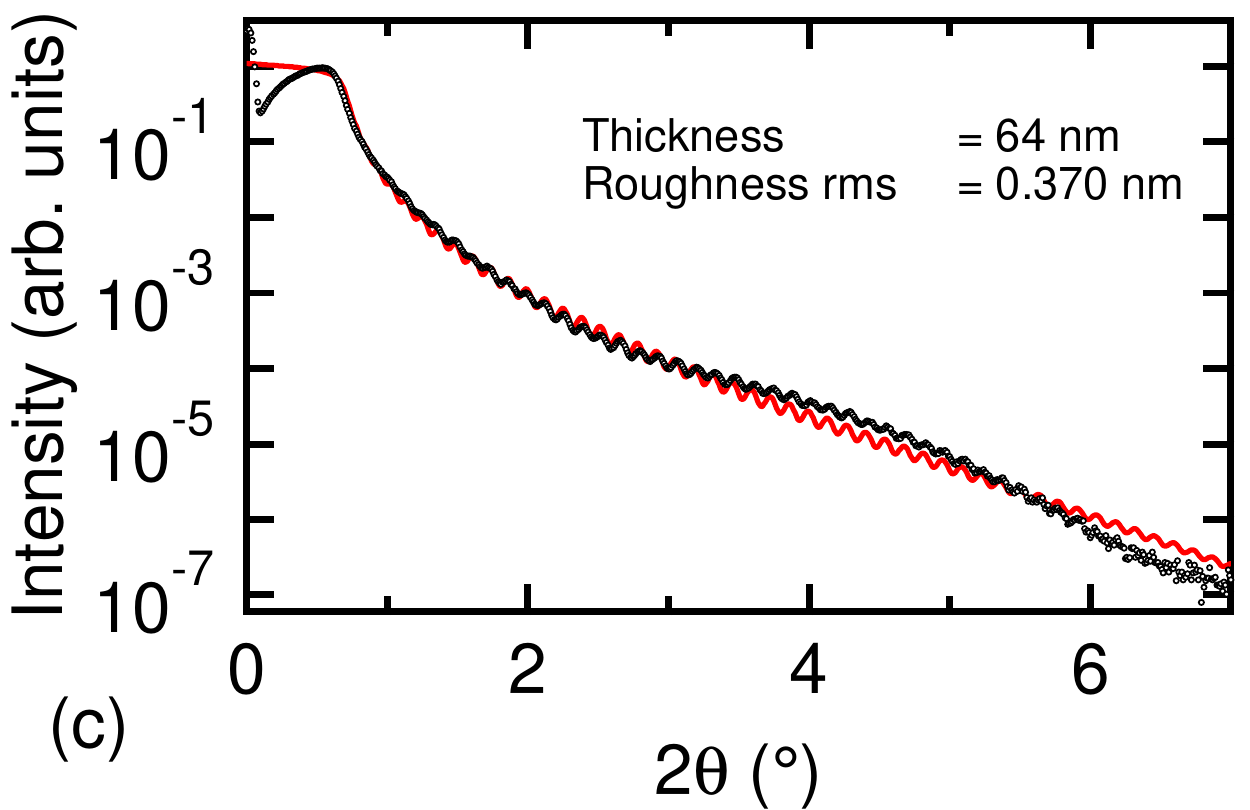}}\hfill
 		\subfloat{\label{fig:LEED}\includegraphics[scale=0.376]{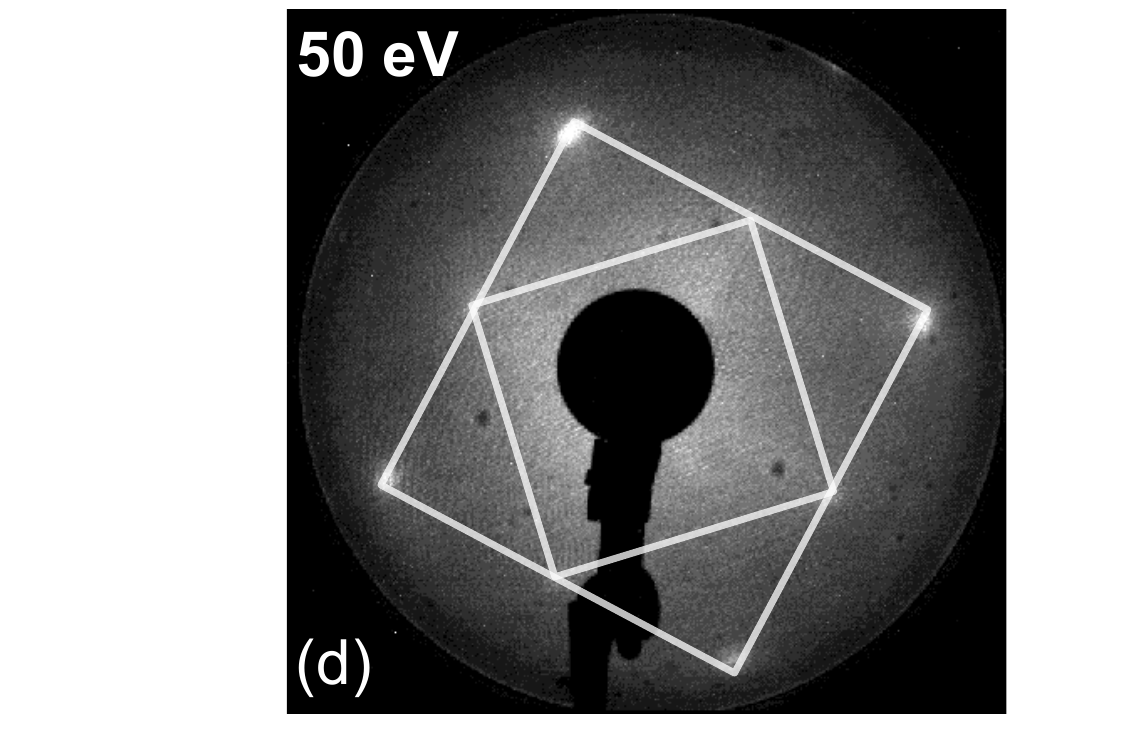}}
   \caption{(002) (a) and (202) (b) peaks for BTO and STO. BTO is relaxed onto STO with a tetragonality ratio c/a of 1.040; (c) reflectivity curve indicating the BTO thickness; (d) LEED image of the BTO surface before Pt deposition showing c(2x2) surface reconstruction}
  \label{fig:carac_Diff}
\end{figure}

The BTO layer is fully relaxed on the NSTO substrate with an in-plane parameter a = 3.956\AA, an out-of-plane parameter c = 4.040~\AA~hence a tetragonality ratio c/a of 1.040 (see Fig.~\ref{fig:XRD_c}-\ref{fig:XRD_a}). The film thickness determined by X-Ray Reflectivity was 64~nm with a 0.37~nm rms roughness (one unit cell) (Fig.~\ref{fig:XRR}). Low Energy Electron Diffraction patterns showed a c(2x2) surface reconstruction (Fig.~\ref{fig:LEED}) expected for TiO$_\mathrm{2}$ termination~\cite{kolpak_evolution_2008}.
	
\begin{figure*}[ht]
	\includegraphics[scale=0.6]{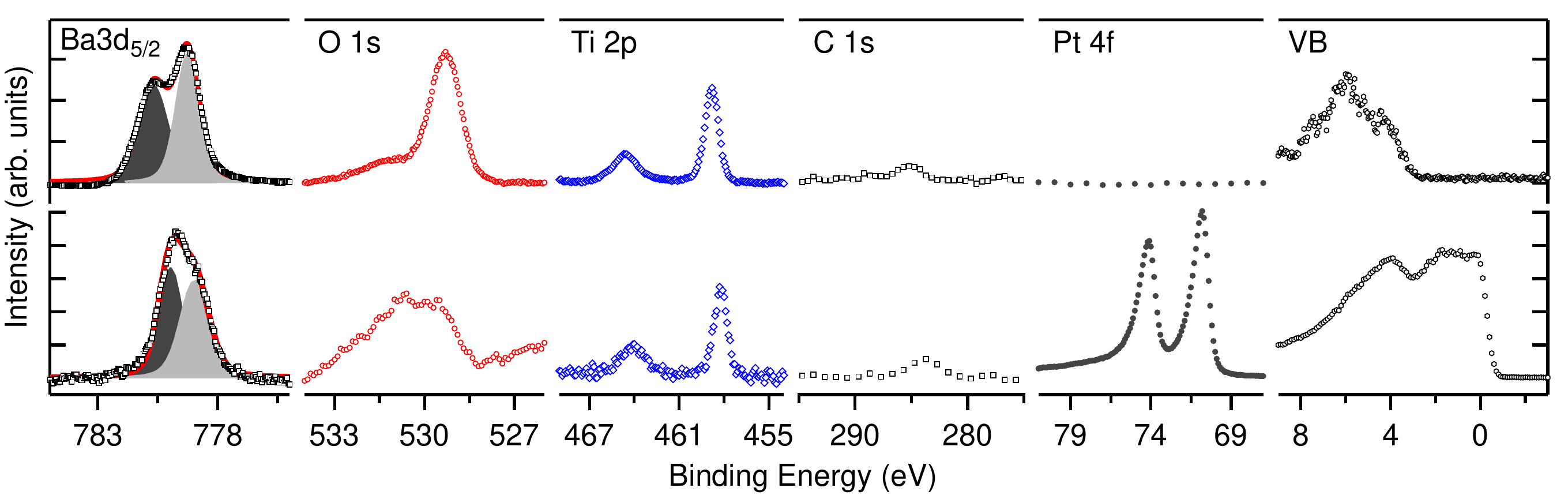}
  \caption{XPS spectra before (top) and after (bottom) Pt deposition.}
  \label{fig:XPS_labo}
\end{figure*}

Fig.~\ref{fig:XPS_labo} shows XPS core-level spectra from the BTO thin film before and after Pt deposition. The secondary electron background has been removed using a Shirley background~\cite{shirley_high-resolution_1972}. Ba~3d and Ti~2p spectra are fitted with Voigt functions. The FWHM of each component of the Ba~3d$_{5/2}$ and Ti~2p$_{3/2}$ are 1.40~eV (Ba~3d$_{5/2}$ HBE), 1.20~eV (Ba 3d$_{5/2}$ LBE) and 1.00~eV (Ti~2p$_{3/2}$). The low-binding-energy (LBE) component is from the bulk-coordinated Ba. The Ba~3d$_{5/2}$ spectrum has a high-binding-energy (HBE) component often attributed to under-coordinated Barium at the surface for BaO termination~\cite{hudson_surface_1993, li_experimental_2005, guo_structural_2012}. However, for a TiO$_\mathrm{2}$ terminated surface, this is erroneous. The topmost Ba atoms are in the first layer below the surface and are fully oxygen coordinated therefore we do not expected a surface peak due to undercoordinated Ba. It is also unlikely that the HBE component is due to contamination. Firstly such peaks (Ba-CO$_\mathrm{3}$ for example) have a much bigger shift~\cite{baniecki_surface_2006} and secondly the Ba is protected from surface contamination by the TiO$_\mathrm{2}$ termination layer. Here, it is more likely caused by the discontinuity in the polarization at the surface for this TiO$_\mathrm{2}$ termination. The discontinuity creates a surface dipole giving rise to an effective core level shift for the first BaO layer (see Refs.~\onlinecite{fechner_effect_2008, wang_chemistry_2012}).This will be discussed further in section~\ref{ssec:response}. O~1s and C~1s spectra before deposition show that there is a very low concentration of carbonates species at the interface confirming the efficiency of ozone cleaning on these surfaces. Ti~2p spectra do not show a shoulder at lower binding energy, classically attributed to Ti$^{3+}$ species due to oxygen vacancies~\cite{schafranek_barrier_2008}, neither before nor after Pt deposition. Additionally, measurements of the valence band of BTO at a take-off angle of 30$^{\circ}$ before deposition showed no signal in the BTO gap which would corroborate the presence of oxygen vacancies at the BTO surface below our detection limit (1\%). The Pt/BTO interface presented here therefore has very low carbonate and oxygen vacancy concentrations. We observe a core-level shift towards lower-binding energy after Pt deposition as previously observed for the (Ba,Sr)TiO$_2$/Pt interface~\cite{schafranek_barrier_2008}. These shifts are classically attributed to a band-bending phenomenon at the metal/ferroelectric interface. In that interpretation, BTO is seen as a n-type, wide-gap semiconductor because of the electrons released by oxygen vacancies. However this model does not fit well with our case in which the core-levels are not rigidly displaced: Ti~2p$_{3/2}$ shifts by 500~meV, the HBE (LBE) component of Ba~3d$_{5/2}$ by 350~meV (700~meV). Furthermore, the oxygen vacancy concentration associated with the n-type conductivity is very low. A surface photo voltage due to the photoemission process might alter the band-bending but in that case all the bands should be rigidly displaced~\cite{hecht_role_1990}. The different core-level shifts can be due to the different chemical environment but also to the change in electrical boundary conditions related to a different screening of the ferroelectric polarization for a BTO free surface and a Pt/BTO interface. This will be further investigated using photoemission spectroscopy with in-situ applied bias in section~\ref{XPS_bias}.

	\subsection{Electrical characterization}
	
\begin{figure}[ht]
  \centering
		\subfloat{\label{fig:dispo_CV}\includegraphics[scale=0.47]{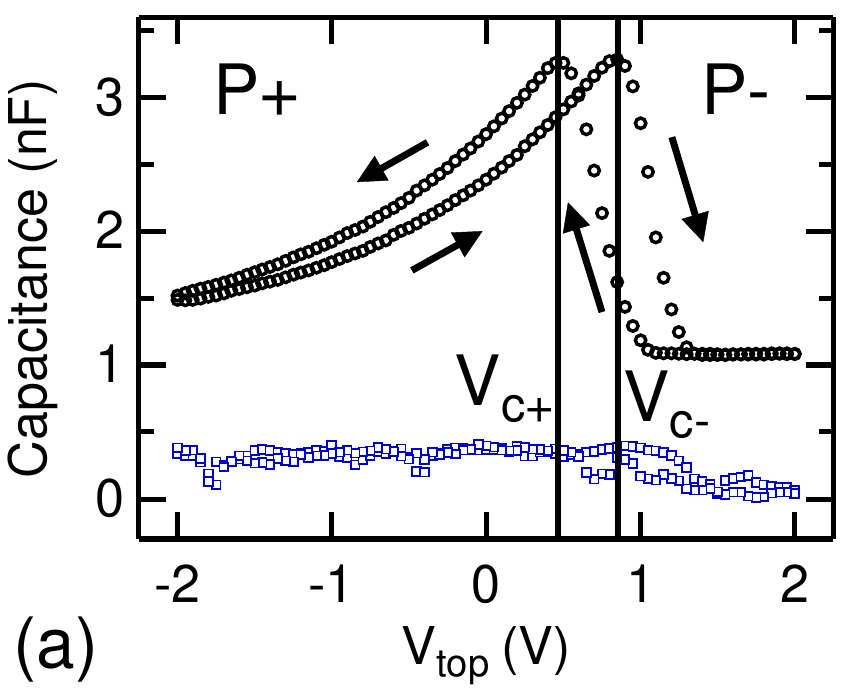}}
		\subfloat{\label{fig:dispo_IV}\includegraphics[scale=0.47]{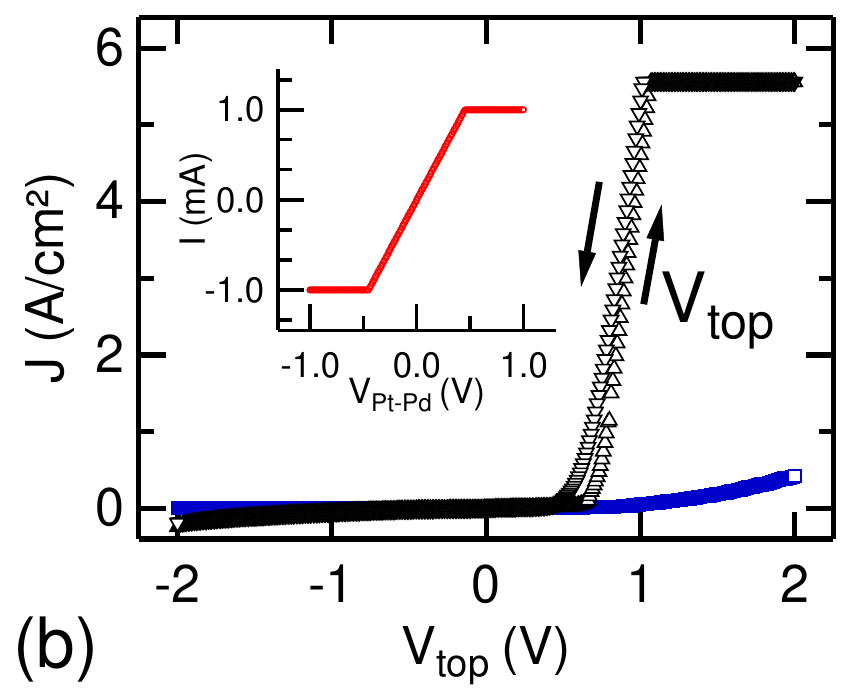}}\\	
 		\subfloat{\label{fig:IV_log}\includegraphics[scale=0.47]{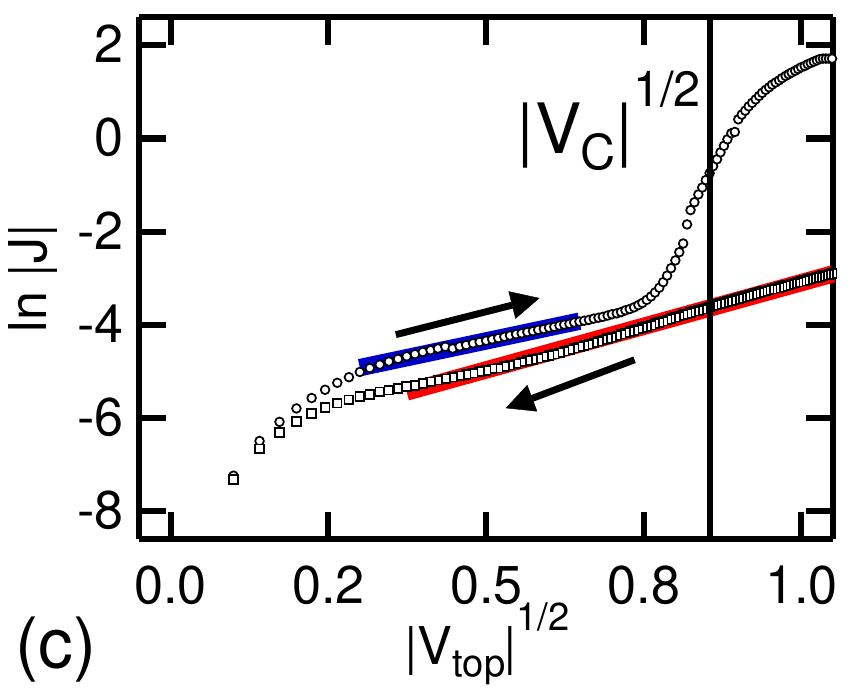}}
 		\subfloat{\label{fig:IV_ohmic}\includegraphics[scale=0.47]{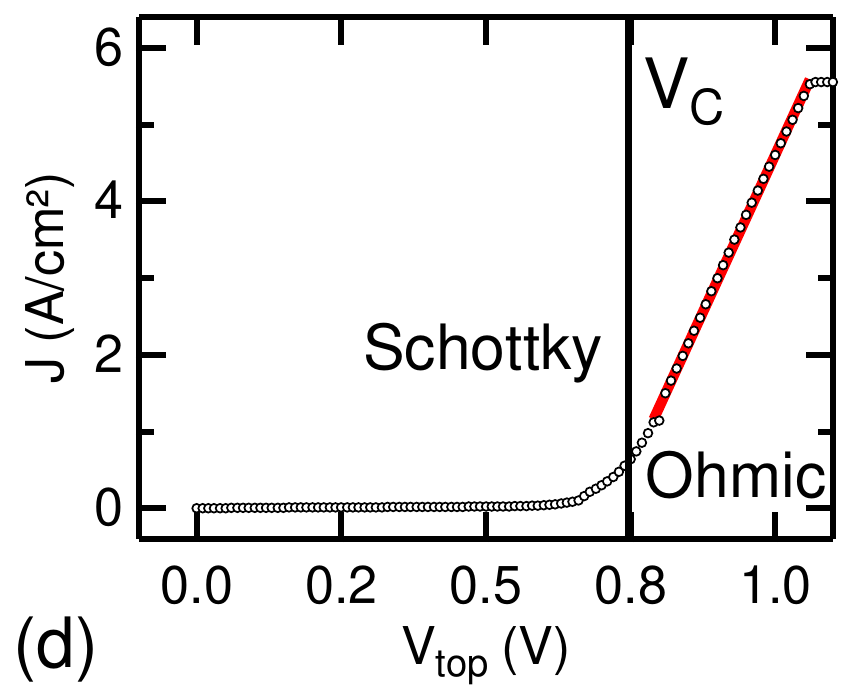}}
   \caption{(a) C-V measurements for Pt/BTO/NSTO (black circles) and Pd/Al$_\mathrm{2}$O$_\mathrm{3}$/BTO/NSTO (blue squares); (b) J-V measurements for Pt/BTO/NSTO (black triangles, upward (downward) triangles for increasing (decreasing) voltage sweep), Pd/Al$_\mathrm{2}$O$_\mathrm{3}$/BTO/NSTO (blue squares) and Pd/Pt (inset).(c) $ \mathrm{ln}~|J| - |V_{top}|^{1/2}$ showing conduction limited by Schottky emission (arrows indicates sweep direction); (d) J – V showing ohmic behavior for V$_\mathrm{top} > V_\mathrm{c}$.}
  \label{fig:dispo}
\end{figure}

After the ex-situ microfabrication processes described in section~\ref{sec:experiment}, the capacitance--voltage (C-V) characteristic (Fig.~\ref{fig:dispo_CV}) has a typical butterfly loop shape as previously observed on Pt/BTO/NSTO~\cite{abe_asymmetric_1997} and Pt/BTO/NSTO~\cite{khassaf_potential_2012} structures. The butterfly loop shows a 0.6~V offset, demonstrating FE behavior with a strong built-in P+ (pointing from the bottom to the top interface) polarization. The coercive voltages V$_\mathrm{c-}$ needed to switch from P+ to P- is 0.80~V. V$_\mathrm{c+}$, switching from P- to P+, is 0.40~V. The accurate measurement of these values is extremely important for application of correct bias voltages in-situ during the XPS measurements. The current density--voltage (J-V) characteristic is diode-like with high conduction when switching from P+ to P-. The characteristics was acquired from P+ to P- and P- to P+ (see Fig.~\ref{fig:dispo_IV}) to confirm that the high conducting state occurs for downward (P-) polarization only. The J-V characteristic was also measured on the Pd pad to check we have an ohmic contact between Pd pads and Pt electrodes and that the Pd/Al$_\mathrm{2}$O$_\mathrm{3}$/BTO/NSTO path is highly insulating ($> 1$~Mohm, blue squares in Fig.~\ref{fig:dispo_IV}). The C-V curve of Pd/Al$_\mathrm{2}$O$_\mathrm{3}$/BTO/NSTO also confirms that parasitic signal from this capacitance is negligible ($< 1$~pF, blue squares in Fig.~\ref{fig:dispo_CV}) in comparison to that from the Pt/BTO/NSTO ($\approx$ 1~nF, black circles in Fig.~\ref{fig:dispo_CV}).

Fig.~\ref{fig:IV_log} displays a plot of $\mathrm{ln}~|J|$ against $|V_{top}|^{1/2}$ obtained by sweeping the applied voltage (V$_\mathrm{top}$) from -2 to +2~V. Starting in the direction of the lower arrow, this corresponds to a switch from P+ to P- polarization while measuring the current through the capacitor. For V$_\mathrm{top} < 0$ (squares in Fig.~\ref{fig:IV_log}), we obtain a good linear fit of $\mathrm{ln}~|J|$ typical of Schottky thermionic emission~\cite{pintilie_ferroelectric_2007}. Similarly, for V$_\mathrm{top} > 0$ (upper branch in Fig.~\ref{fig:IV_log}), we obtain a linear fit but for a smaller voltage range: $0.2 < |V_{top}|^{1/2} < 0.8~ \mathrm{volt}^{1/2}$. In Fig.~\ref{fig:IV_ohmic}, for V$_\mathrm{top} > V_\mathrm{c} = 0.8~\mathrm{volt}$, the linear fit to the experimental curve shows an ohmic response. V$_\mathrm{c}$ matches the coercive voltage deduced from the C-V measurements. Therefore, when switching from P+ to P-, we observe a transition from Schottky-like to ohmic conduction. Since high-flux high-energy incident photons might affect the ferroelectric and conduction properties of the heterostructure~\cite{wu_direct_2011}, we did C-V measurements under X-ray illumination, these are discussed in the following section.

	\subsection{Photoelectron Spectroscopy with applied bias}
	\label{XPS_bias}

To understand the microscopic electronic structure responsible for the electrical properties, photoelectron spectroscopy was carried out on the heterostructure at different values of applied bias on the top (Pt) electrode or bottom (NSTO) electrodes. 
	
\begin{figure}[ht]
  \centering
  		\subfloat{\label{fig:CV_photons}\includegraphics[scale=0.45]{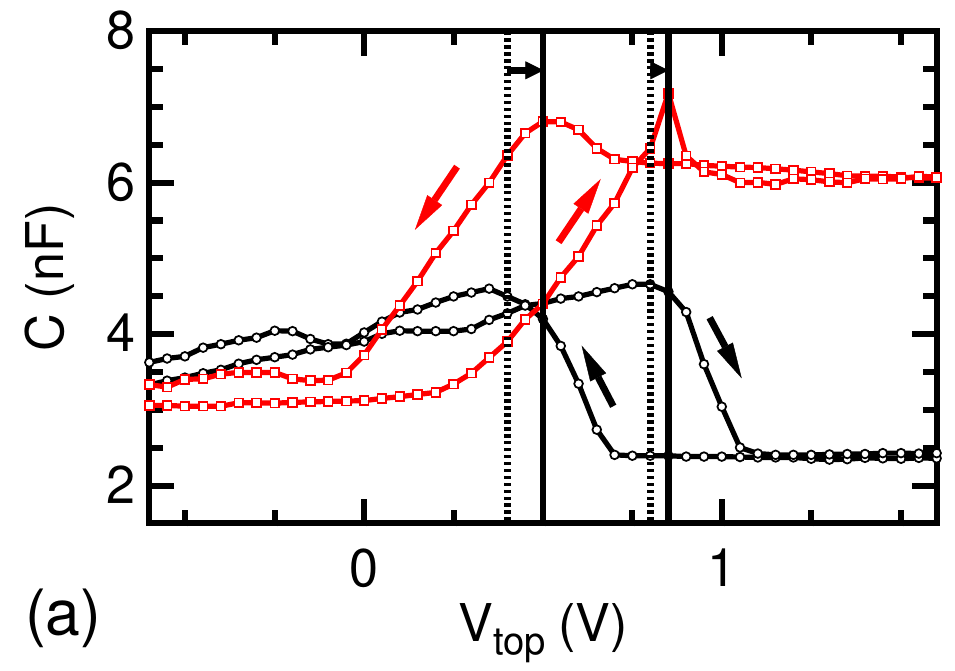}}
		\subfloat{\label{fig:photons_theory}\includegraphics[scale=0.45]{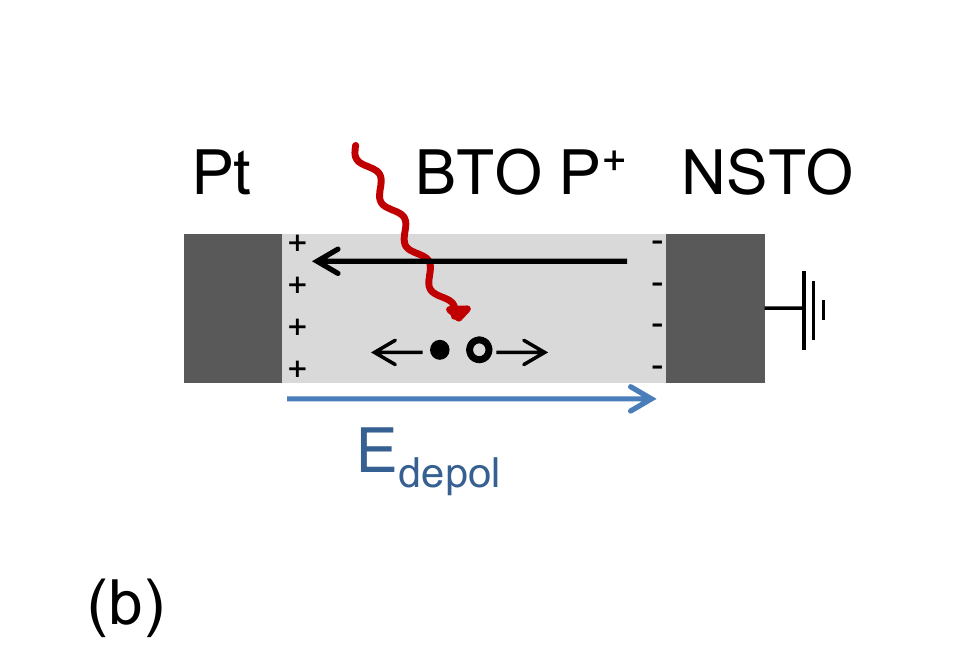}}
  \caption{(a) Capacitance without C$_\mathrm{dark}$ (black circles) and with C$_\mathrm{photon}$ (red squares) synchrotron irradiation,(b) capacitor under illumination, short-circuit. Photogenerated electrons (holes) are filled (empty) circles.}
  \label{fig:in_situ_elec}
\end{figure}	
	
Fig.~\ref{fig:CV_photons} displays the C-V measurements with and without synchrotron radiation on the Pt top electrode. It shows that the coercive voltages change slightly when the sample is illuminated: V$_\mathrm{c+}$ is shifted up from 0.40 to 0.50~volt, V$_\mathrm{c-}$ shifts up from 0.80 to 0.85~volt. The alteration of the C-V curve only occurs when the photon beam is focused on the platinum top electrode: no change is observed when focusing on the Pd pad or nearby free BTO. In the BTO layer, photo-generated carriers drift under the effect of the depolarizing field and the applied bias. For the P+ state for instance, electrons (holes) migrates towards top (bottom) interface and contribute to a better screening of the depolarizing field induced by the P+ polarization. This might lead to a higher stability of the P+ state, and an overall shift of the C-V loop towards higher voltages. Though weak, this effect was monitored for every photoemission measurements: before each set of acquisition, CV loops were acquired to check the coercive voltages.

\begin{figure}
  \centering
  	\includegraphics[scale=0.53]{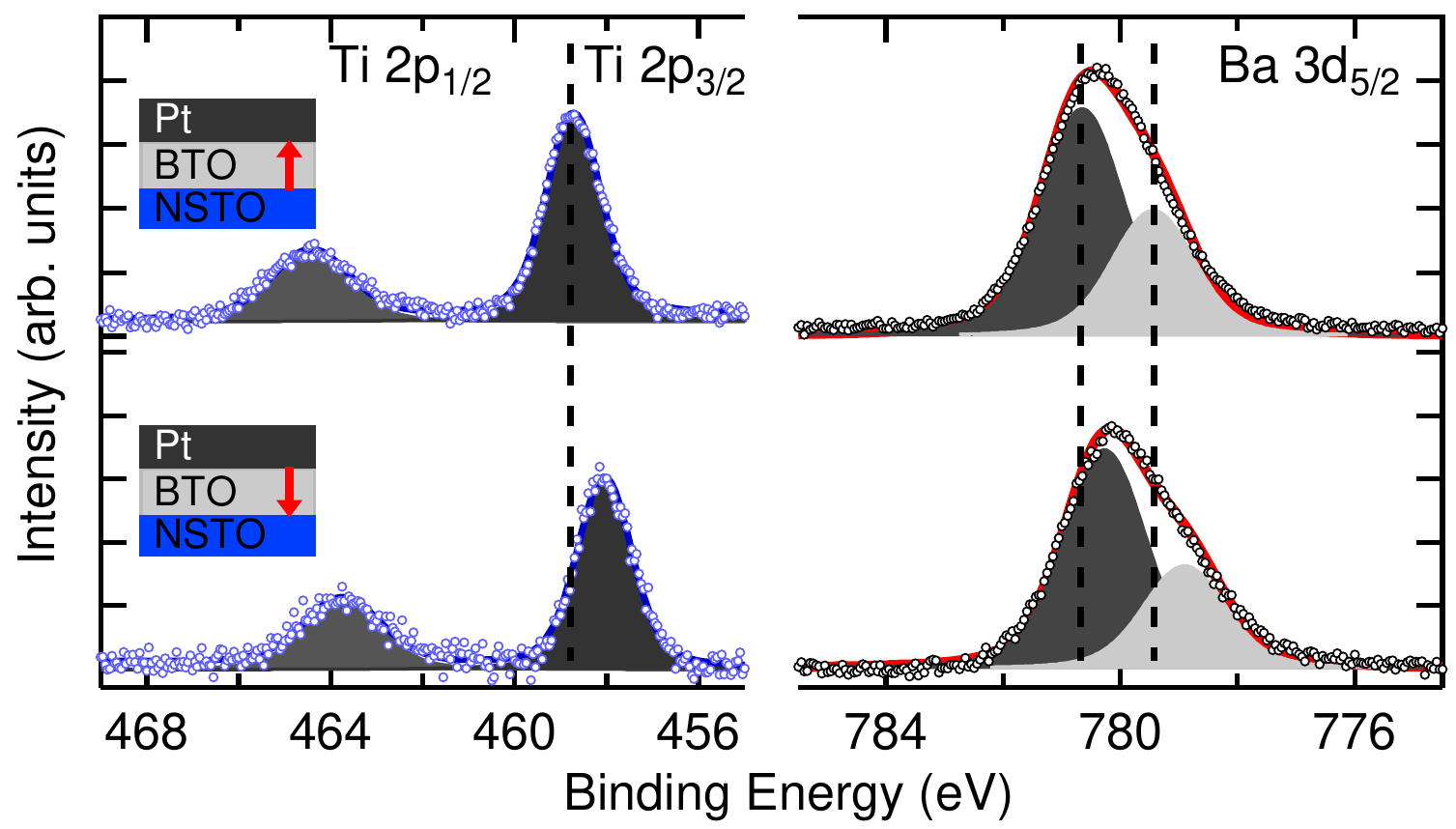}
  \caption{Photoemission spectra for the P+ (top) and P- (bottom) FE states for the Ti $2p_{3/2}$ - $2p_{1/2}$ (left) and Ba $3d_{5/2}$ (right). The bottom spectra have been shifted by 0.70~eV towards lower BE using Pt~4f spectra.}
  \label{fig:XPS_TEMPO}
\end{figure}

The bias-induced displacements of Ba~3d and Ti~2p core levels binding energies (BE) were measured relative to the Pt~4f core-levels of the top electrode (see Fig~\ref{fig:XPS_TEMPO}). The Ba~3d$_{5/2}$ spectra again have two components: at low and high binding energy (LBE and HBE respectively). The spectra were fitted using Voigt functions with FWHM of 1.60~eV (Ba~3d$_{5/2}$) and 1.40~eV (Ti~2p). For the P+ (P-) polarization state, we kept V$_\mathrm{top}$ at -0.5~V (+1.30~V) \textit{i.e.} well above the coercive voltages, for two minutes to ensure full switching of the polarization before recording the photoemission spectra at V$_\mathrm{top}$ = 0~V (+0.75~V). The lower voltage magnitudes are chosen to reduce the current flowing through the capacitor while maintaining the correct polarization state. The Pt~4f core-level shifts towards higher BE is 0.70~eV when we apply +0.75~V on the Pd pad. The 0.05~V potential drop is due to a small contact resistance between the two metals ($\approx 10 \Omega$). When switching from P+ to P- state, the Ba~3d$_{5/2}$ HBE (LBE) shifts downward by 200~meV (300~meV) and the Ti~2p$_{3/2}$ by 450~meV relative to the Pt~4f$_{7/2}$ reference (Table~\ref{tab:BE_BTO_top}) The last column in Table~\ref{tab:BE_BTO_top} shows the BE shifts for BTO core-levels after subtraction of the applied bias. These values are therefore the direct consequence of switching the FE polarization. 

We did the same experiment applying bias on the bottom electrode with top electrode grounded. The measured BE shifts are equal (within 50~meV) to those with voltage on top electrode (see Table~\ref{tab:BE_BTO_bottom}).

\begin{table}
	\begin{tabular}{|l|c|c||c|}  
	\hline
	\multirow{2}{*}{Core-Level} & \multicolumn{2}{c||}{V$_\mathrm{top}$}&  \\
								& \quad 0.00 (P+) & \quad 0.75 (P-) & $\Delta_{BE}$ wrt Pt~4f \\ \hline	
  Pt~4f$_{7/2}$ & 71.20 & 71.90 & 0.00 \\
  \hline
  Ba~3d$_{5/2}$ LBE & 779.45 & 779.80 & -0.35 \\
  \hline
  Ba~3d$_{5/2}$ HBE & 780.65 & 781.15 & -0.20 \\
  \hline
  Ti~2p$_{3/2}$ & 458.70 & 458.95 & -0.45 \\
  \hline
	\end{tabular}
\caption{Binding energy (eV) for the Pt/BTO interface core-levels when applying different biases V$_\mathrm{top}$ (V) (column 1,2) and shift with respect to Pt~4f$_{7/2}$ BE shift (column 3). Bottom electrode is grounded.}
\label{tab:BE_BTO_top}
\end{table}

\begin{table}
	\begin{tabular}{|l|c|c||c|}  
	\hline
	\multirow{2}{*}{Core-Level} & \multicolumn{2}{c||}{V$_\mathrm{bottom}$} &  \\
								& \quad 0.00 (P+) & \quad -0.95 (P-) & $\Delta_{BE}$ wrt Pt~4f \\ \hline		
  Pt~4f$_{7/2}$ & 71.20 & 71.20 & 0.00 \\
  \hline
  Ba~3d$_{5/2}$ LBE & 779.60 & 779.30 & -0.30 \\
  \hline
  Ba~3d$_{5/2}$ HBE & 780.70 & 780.55 & -0.15 \\
  \hline
  Ti~2p$_{3/2}$ & 458.75 & 458.30 & -0.45 \\
  \hline
	\end{tabular}
\caption{Binding energy (eV) for the Pt/BTO interface core-levels when applying different biases V$_\mathrm{bottom}$ (V) (column 1,2) and BE offsets with respect to Pt~4f$_{7/2}$ BE shift (column 3). Top electrode is grounded.}
\label{tab:BE_BTO_bottom}
\end{table}

The valence band offset (VBO) at the Pt/BTO interface may be determined as follows~\cite{kraut_precise_1980}:

\begin{eqnarray}
\label{eq:VBO}
\mathrm{VBO} =& (E_{Ti2p_{3/2}} – E_{Pt4f_{7/2}})_{Pt/BTO} \\ &+ (E_{Pt4f_{7/2}} – E_F)_{Pt} - (E_{Ti2p_{3/2}} – \mathrm{\textit{VB}}_\mathrm{Max})_{BTO} \notag
\end{eqnarray}

$(E_{Ti2p_{3/2}} – E_{Pt4f_{7/2}})_{Pt/BTO}$ is extracted from the \textit{in-situ} photoemission measurements (see Table~\ref{tab:BE_BTO_top}). $(E_{Pt4f_{7/2}} – E_F)_{Pt} = 71.25~eV$ was measured on a Pt(111) monocristal and $(E_{Ti2p_{3/2}} – VB_\mathrm{Max})_{BTO} = 455.95~\mathrm{eV}$ on the clean BTO thin film before Pt deposition, giving a valence band offset of 2.8~eV. Assuming that the BTO thin film has the same gap as in the bulk (3.2~eV~\cite{wemple_polarization_1970}) and that it does not change with polarization, we deduce a 0.4~eV (0.85~eV) conduction band offset (CBO) for the P+ (P-) case. 

\section{Discussion}
	\subsection{Polarization-dependent band lineup}

\begin{figure}[ht]
  \centering
  	\includegraphics[scale=0.53]{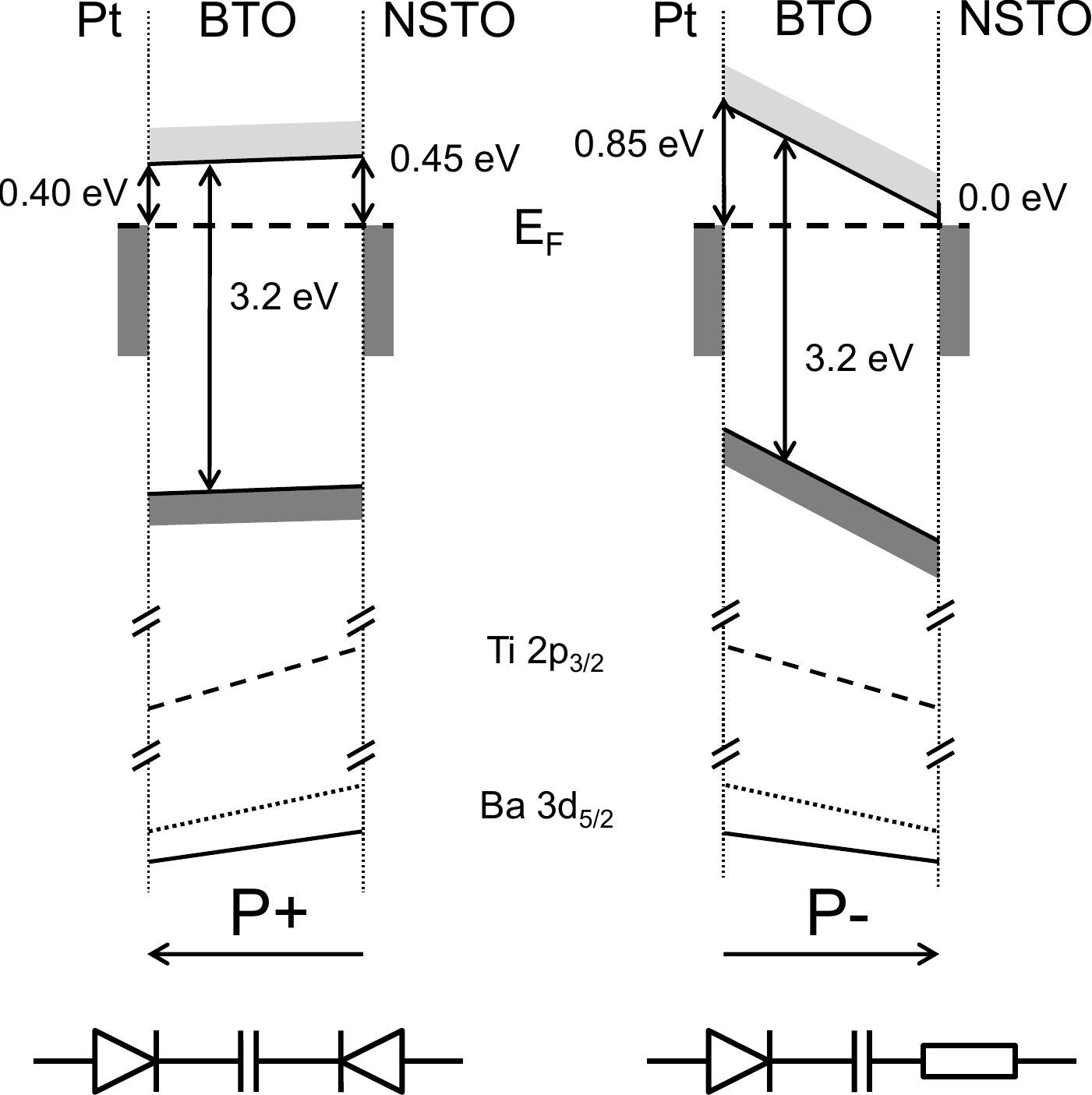}
  \caption{Band lineup for the P+ (left) and P- (right) polarization states.}
  \label{fig:band_lineup}
\end{figure}

The inelastic mean free path of photoemitted electrons by soft x-ray photons is of 3-4~nm~\cite{brundle_elucidation_1975} limiting the probing depth of this experiment to the first layers of BTO beneath the platinum overlayer. The BTO in the center of the film or at the bottom interface is not probed. Hard x-ray photoemission spectroscopy experiments using photon energy of 6-10~keV cannot probe more than 20-30~nm~\cite{panaccione_hard_2012}. Therefore, the signal from the deeply buried bottom interface of our film cannot be directly probed with photoemission experiment. We can deduce the barrier height at the bottom interface from electrical measurements. For both P+ and P- polarization states, the barrier heights at the top interface (0.40 and 0.85~eV) are large enough to suggest Schottky conduction. For negative V$_\mathrm{top}$ a Schottky-like behavior is observed for the whole capacitor. Therefore the bottom interface barrier height must be at least similar to that of Pt/BTO. For positive V$_\mathrm{top}$, the Pt/BTO barrier height increases, therefore in order to have the observed ohmic behavior, the BTO/NSTO barrier height must be very small. More specifically, the ohmic conduction for V$_\mathrm{top} > V_\mathrm{c} = 0.8~\mathrm{volt}$ suggests a conduction band offset of 0~eV at the bottom interface for this voltage range (P- state). When $V_\mathrm{top} < V_\mathrm{c}$ (Fig.~\ref{fig:band_lineup}, right), the BTO layer switches from P- to P+. Using photoemission spectroscopy data we obtain a downward shift of 0.45~eV in the conduction band offset at the top interface. The offset is now 0.85~eV at the Pt/BTO interface. Assuming a upward shift of similar magnitude at the bottom interface, we get a conduction band offset value of 0.45~eV for the bottom interface in the P+ state. However, this analysis must be treated with caution since a higher upward shift might occur at the bottom interface~\cite{umeno_ab_2009}. 

The full band alignment is shown in Fig.~\ref{fig:band_lineup}. In the P+ state, the structure can be modeled as back-to-back Schottky diodes. The conduction band offset is 0.40~eV at the Pt/BTO interface, similar to the value obtained in Ref.~\onlinecite{schafranek_barrier_2008} for (Ba,Sr)TiO$_\mathrm{3}$/Pt (0.5~eV) and 0.45~eV at the BTO/NSTO interface. For V$_\mathrm{top} < 0$, the Pt/BTO diode is reverse-biased while the BTO/NSTO junction is forward biased. For $0 < V_\mathrm{top} < V_\mathrm{c}$, the reverse-biased BTO/NSTO diode limits the electron current.

In the P- state, the BTO/NSTO interface is ohmic and the structure can be modeled as a unique Schottky-diode. This diode is forward-biased when $V_\mathrm{top} > 0$, leading to relatively high current through the capacitor. The whole structure therefore functions as a resistance switch with a bistable state at a non-zero voltage. An electroresistance can even be defined, in an analogous way to the tunneling electrical resistance, as $ (G_{-} - G_{+}) / (G_{-} + G_{+})$. We extracted the current for a voltage of 0.65~V on both sides of the I-V characteristic, i.e. coming from P+ and from P- and obtained an electroresistance of 0.79.

The observed band offsets can be compared to the predictions based on the theory of metal-semiconductor junctions extended to ferroelectric materials by Pintilie \textit{et al.} in Ref.~\onlinecite{pintilie_ferroelectric_2007}. Using Equation~(11) from this reference, the theoretical magnitude of barrier height change due to the FE polarization in comparison to a virtual paralectric (PE) state is:

\begin{eqnarray}
\label{eq:polar_theory}
\mathrm{CBO}_\mathrm{FE} – \mathrm{CBO}_\mathrm{PE}= \sqrt{\frac{eP}{4 \pi \epsilon^2_0 \epsilon_\mathrm{opt} \epsilon_r }}
\end{eqnarray}

P = 0.27~C/m$^2$ ~\cite{wieder_electrical_1955} and the optical permittivity is $\epsilon_\mathrm{opt}$ = 5.40 extrapolated from refractive index measurements in Ref.~\onlinecite{ghosez_lattice_1999}. At voltage far from the coercive voltages, capacitance tends to $\epsilon_r \epsilon_0 \frac{A}{d}$, where A = 300x300~$\mu$m$^\mathrm{2}$ is the area of the electrode and d = 64~nm is the BTO layer thickness. From Fig.~\ref{fig:dispo_CV}, we estimate the dielectric permittivity $\epsilon_r \approx 90$. Using these values in equation~(\ref{eq:polar_theory}), FE polarization induces a decrease (increase) of 0.30~eV of the top interface conduction band offset in the P+ (P-) state compared to the PE state. The overall upward 0.60~eV shift when switching from P+ to P- is comparable with our spectroscopic measurements.

Fig.~\ref{fig:band_lineup} shows that the P- state leads to a highly asymmetric band alignment with a 0.85~eV (0~eV) top (bottom) conduction band offset. The combination of asymmetric electrodes and P- (P+) FE polarization lead to an increased (decreased) slope of the BTO conduction band. Consequently, BTO in the P- (P+) state experiences an enhanced (reduced) depolarizing field $E_\mathrm{depol}$ in comparison to a symmetric capacitor. The internal field can be estimated from the slope of the CB over the whole BTO layer. We obtain $\mathrm{E}_\mathrm{depol} = 130~\mathrm{kV/cm}$ for the P- state. Using the mean coercive voltage $ (V_{c-} - V_{c+}) / 2$, we estimate the coercive field to be about 30 kV/cm: a hypothetical P- state will always tend to switch back to P+ in short-circuit boundary conditions (zero voltage). FE instability due to asymmetric interfaces has recently been reported for SRO/BTO/SRO capacitors where the bottom SrO/TiO$_\mathrm{2}$ and top BaO/RuO$_\mathrm{2}$ make the P+ state unstable~\cite{lu_enhancement_2012}.

In Ref.~\onlinecite{chen_polarization_2012}, Chen and Klein measure a conduction band offset of 0.30~eV (0.95~eV) at the interface between a thin layer of platinum and a single crystal of BTO in the P+ (P-) state. Polarization switching induces a 0.65~eV barrier height shift, to be compared to 0.45~eV in our case. The difference cannot only be due to different energy resolution. Our band offset calculations are the same as Chen and Klein, and $(E_{Ti2p_{3/2}} – VB_\mathrm{Max})_{BTO}$ was almost identical in both experiments. For a given material with similar polarization magnitude, a lower barrier shift can be due to a better screening of the ferroelectric polarization, i.e.\ to a smaller screening length. The flux delivered by an undulator in a synchrotron facility is significantly higher than from a laboratory x-ray source and might be responsible for a part of the enhanced screening. However, the data of section~\ref{XPS_bias} show a rather weak effect of the photon-induced screening. More likely, the interface configuration is responsible for the different screening. As pointed out by Rao \textit{et al.}, the Pt/BTO interface shows multiple configurations with different screening capabilities~\cite{rao_structural_1997}. The two heterostructure are different and this can explain the different barrier heights. The ferroelectric layer is also not the same. The BTO single crystal studied by Chen and Klein is significantly thicker (0.4~mm) than our BTO thin film (64~nm) leading to a lower depolarizing field hence different interface screening. Their single crystal surface termination and reconstruction are not known and might be different from ours. Finally, the full electrode/ferroelectric/electrode structure is different. They used a symmetric Pt/BTO/Pt capacitor compared to the asymmetric Pt/BTO/NSTO structure presented in this article. The asymmetry has a direct influence on the internal field inside the ferroelectric layer as shown in Fig~\ref{fig:band_lineup}. All of these differences can account for the barrier height discrepancies between the two studies. More systematic studies are required in order to identify the respective contributions.

	\subsection{Electronic bands response to applied bias}
	\label{ssec:response}

Although the barrier heights agree with Ref~\onlinecite{pintilie_ferroelectric_2007}, interestingly, switching the polarization does not affect all the core levels in the same way. This is extremely important because much analysis of band alignment at a FE/electrode interface assumes rigid band shifts, as in classical semiconductor physics as done by Kraut \textit{et al.} in Ref.~\onlinecite{kraut_precise_1980} and also underpins equation~(\ref{eq:VBO}). However, the Ti~2p core level is shifted by 450~meV compared with 350~meV (200~meV) for the LBE (HBE) Ba~$3d_{5/2}$. A rigid shift in the BTO band structure due to the polarization-induced interface dipole cannot explain the different shifts of the Ti and Ba core-levels~\cite{wu_direct_2011}. This is unexpected and has not been previously reported in the literature. Our group has observed such behavior on BTO single crystals using spatially-resolved photoemission spectroscopy on oppositely-polarized domains (unpublished results). 

First principles calculations~\cite{stengel_band_2011} showed that near the top electrode of an SRO/BTO/SRO heterostructure in a P+ polarization state, the Ti~3d conduction band cross the Fermi level, allowing charge spillage into the BTO~\cite{stengel_band_2011}. A P+/P- shift of 1.4~eV in the band structure is predicted, with the conduction band next to the top electrode in the P+ state 0.3~eV below the Fermi level. This so-called pathological case of band bending is due to the LDA band gap in these calculations of 1.8~eV. However, if we take the experimental band gap of 3.2~eV and assume that both the VBM and the conduction band offset are increased by about the same amount (0.7~eV) with respect to the LDA values, then the experimental conduction band offset should be 0.4~eV above the Fermi level, which is precisely what we measure. Such a good agreement is most probably fortuitous, for example, the higher screening by Pt with respect to SRO used by Stengel et al. might modify this value. The P+/P- shift in the band positions will certainly change as a function of the screening efficiency of the electrode material, which explains why we measure a shift of 0.45~eV compared to the theoretical prediction of 1.4~eV. Nevertheless, the trend is consistent with the interpretation that the origin of the band alignment lies in the imperfect screening by the electrode creating a voltage drop and the residual depolarizing field inside the FE, altering the electrostatic potential. Yet, this still does not account for the differences in the experimentally measured P+/P- Ba and Ti core level shifts.

To do so, we tentatively propose an explanation based on dynamical charge in the vicinity of the interface, i.e. the change in polarization created by an atomic displacement, which is a more measurable quantity. By describing the changes in orbital population as a function of atomic displacements, dynamical charge gives a direct picture of changes in covalency/ionicity~\cite{ghosez_dynamical_1998}. The relevant quantity is the Callen (or longitudinal) charge since we apply a non-zero external field to the BTO layer. It is related to the transverse or Born effective charge by the optical dielectric tensor~\cite{martin_direct_1981}. DFT simulations on the Pt/BTO interface calculated the lattice mediated contribution to the static inverse capacitance in terms of the Callen dynamical charge and the longitudinal force constant~\cite{stengel_enhancement_2009}. The latter has a short range interaction and creates an enhanced response at the interface. The Callen charge of Ti is twice as high as that of Ba. We suggest that higher dynamical charge may result in a larger core level shift in the FE state with respect to the PE state. Changes in the local atomic distortions due to the proximity of the electrode interface could magnify or reduce this effect. For TiO$_\mathrm{2}$-terminated PbTiO$_\mathrm{3}$ surfaces, Fechner \textit{et al.} showed that the layer polarization near the surface (the two first TiO$_\mathrm{2}$ layers and the first PbO layer) is reduced in comparison to bulk layers~\citep{fechner_effect_2008}. In our experiment, the Ba~3d HBE shift (200~meV), representative of the BaO layer near the interface is indeed smaller than the LBE shift (350~meV), more sensitive to BaO layers deeper in the film. Velev \textit{et al.} investigated Pt/TiO$_\mathrm{2}$-terminated BaTi$_\mathrm{3}$/Pt tunnel junctions using first-principles calculations~\cite{velev_effect_2007}. The Ti displacement at the interface layer in the P- state is close to the centrosymmetric PE position, whereas in the P+ state, the Ti displacement is enhanced relative to bulk FE value. This asymmetry in the atomic displacements should translate in an additional binding energy shift between the corresponding photoemission spectra when switching from P- to P+. The effect is greater for Ti because it is at the interface where the asymmetry in atomic positions is a maximum. Ti also has a larger dynamical charge than Ba, thus the effect of asymmetric atomic displacements on the BE should be amplified. Evidence of similar phenomena can be seen in High-Resolution Transmission Electron Microscopy (HRTEM) measurements on the BiFeO$_\mathrm{3}$/(La, Sr)MnO$_\mathrm{3}$ (BFO/LSMO) interface by Chang \textit{et al.}~\cite{chang_atomically_2011}. For BFO the polarization is dominated by the relative displacement of Fe with respect to Bi. Their results show that on approaching the interface, the layer polarization in BFO is progressively reduced to zero over 4-5 atomic layers for polarization pointing away from the LSMO into the BFO (P-). For polarization pointing towards the LSMO (P+) it remains constant. Although the ferroelectric induced atomic distortions are more complex in BFO than in BTO (rotation and tilting of oxygen octahedra) and the electrode material is not the same, it appears that in the vicinity of the interface the atomic displacements are strongly modified with respect to those in the bulk FE state.

We may therefore have a model which goes beyond the often used non-interacting electron semiconductor vision and which at least qualitatively might account for differential shifts of electron bands under applied bias near the interface. However, this remains hypothetical and further work is necessary, particularly to measure experimentally the layer dependence of the response of the electronic bands to applied voltages. First principles calculations of the Callen charge variations near the interface \emph{under bias} would also be extremely useful. Far from the interface, the dynamical charge for the two polarization states should of course be the same. In addition, we hope that this experimental work will encourage further theoretical work to accurately relate the concept of dynamical charge and core-level binding energies measured by photoemission spectroscopy. Photoemission spectroscopy with bias would be a powerful tool for measuring these quantities.

\subsection{The role of oxygen vacancies}

\begin{figure}[ht]
  \centering
		\includegraphics[scale=0.64]{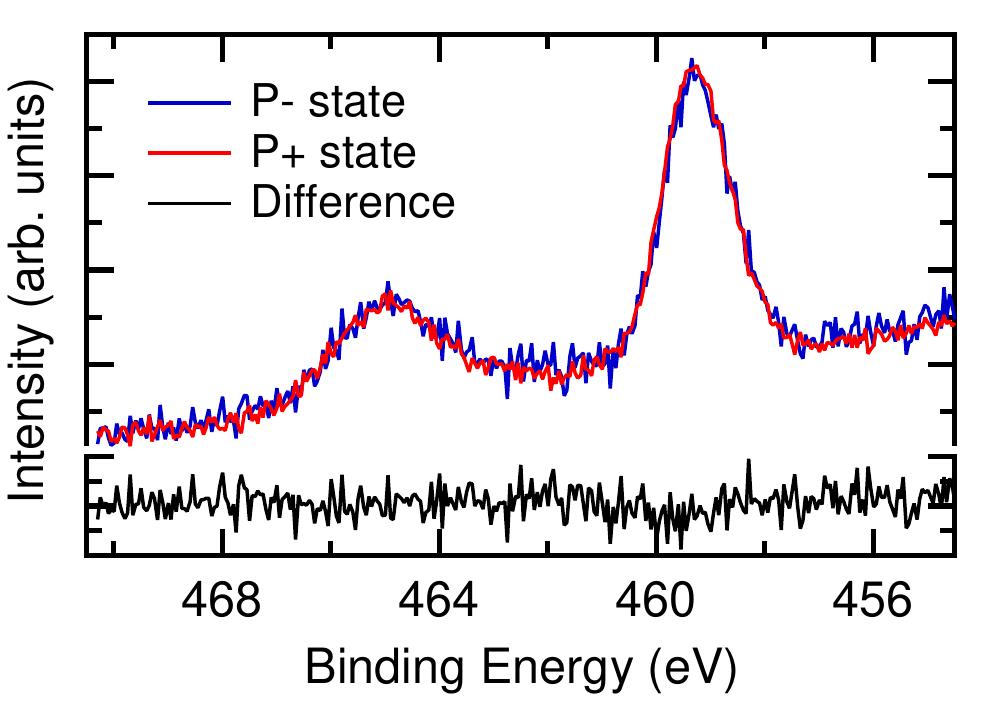}
  \caption{Ti~2p spectra for P+/P- and their intensity difference. P- spectra has been shifted in binding energy using $\Delta_{BE}$ from Table~\ref{tab:BE_BTO_top}. The two spectra have identical shape.}
  \label{fig:vacancies}
\end{figure}

In transition-metal oxides, oxygen vacancies can migrate under the effect of an electric field and drastically changes the barrier properties of the interfaces. This phenomenon can be the cause of resistive switching behavior~\cite{waser_nanoionics-based_2007} similar to the one we measure. Using electrical measurements, Zhang \textit{et al.} explained the diode-like behavior of a Pt/BTO/NSTO heterostructure by oxygen vacancy migration when applying a bias~\cite{zhang_large_2011}. Here, just as before application of bias, there is no LBE shoulder in the Ti~2p spectra typical of Ti$^{3+}$ due to oxygen vacancies. This can be seen from Fig.~\ref{fig:vacancies} which shows P+ and P- Ti~2p spectra, after correction for the polarization-induced BE shifts $\Delta_{BE}$ taken from Table~\ref{tab:BE_BTO_top}. The spectra are identical showing that migration of bulk oxygen vacancies to one interface due to the applied electric field is unlikely and cannot explain the conduction changes, contrary to the conclusions of Ref.~\onlinecite{zhang_large_2011}. However, the latter used thinner films, thus higher applied fields under bias. Their films also had an rms roughness of 5~nm for a 20~nm film, whereas ours show single unit cell steps and flat terrace so are expected to have a much lower defect concentration.

\section{Conclusion}

Photoemission spectroscopy with in-situ bias application has directly measured the band alignment and the electronic structure of FE BTO near the Pt/BTO interface in a Pt/BTO/NSTO heterostructure. The band alignment is determined by a combination of imperfect screening by the electrode and the chemistry of the interface. Callen dynamical charge may offer a plausible explanation of the polarization dependent electronic structure of BTO near the electrode. For a P+ polarization state, i.e. polarization pointing from the bottom electrode to the top electrode, the leakage current is limited by Schottky emission at the Pt/BTO interface whereas in the P- state the structure works as a single Schottky diode with a quasi-ohmic BTO/NSTO interface.

\begin{acknowledgments}
J.R. is funded by a CEA Ph.D. Grant CFR. This work, partly realized on Nanolyon platform, was supported by ANR project Surf-FER, ANR-10-BLAN-1012 and Minos, ANR-07-BLAN-0312. We acknowledge SOLEIL for provision of synchrotron radiation facilities. We thank J. Leroy, S. Foucquart and C. Chauvet for technical assistance.
\end{acknowledgments}

\bibliography{./biblio_BTO_NSTO}

\end{document}